\documentclass[11pt]{article}    
\usepackage[top=1in,bottom=1in,left=1in,right=1in]{geometry}    
\usepackage{diagbox}
\usepackage{multirow}
\usepackage{hyperref}
\usepackage[all]{hypcap}
\usepackage{graphicx}
\usepackage{setspace}
\usepackage[parfill]{parskip}   
\usepackage{amsmath}
\usepackage{longtable}
\usepackage[usenames,dvipsnames]{color}
\usepackage{lscape}
\usepackage[toc,page]{appendix}
\usepackage{amssymb}
\usepackage{lineno}
\usepackage{fancyhdr}
\usepackage{lastpage}
\usepackage{arydshln}
\usepackage[font=small,labelfont=bf]{caption}
\usepackage{authblk}
\usepackage{soul}
\usepackage{lscape}
\usepackage{subfig}
\usepackage{floatrow}
\usepackage{pdflscape}
\usepackage{chapterbib}  
\usepackage{cite}
\usepackage{arydshln}
\usepackage{mathptmx}

\definecolor{commentblue}{rgb}{0.36, 0.40, 0.90}

\title{\textsc{Reflection on modern methods: Good practices for applied statistical learning in epidemiology}}
\author[1,$\dagger$]{Yanelli Nunez}
\author[1,$\dagger$]{Elizabeth A. Gibson}
\author[2]{Eva M. Tanner}
\author[2]{Chris Gennings}
\author[3]{Brent A. Coull}
\author[4]{Jeff A. Goldsmith}
\author[1,$\star$]{Marianthi-Anna Kioumourtzoglou}

\affil[$\dagger$]{Joint first authors}
\affil[1]{Department of Environmental Health Sciences, Columbia University Mailman School of Public Health, New York, NY}
\affil[2]{Department of Environmental Medicine and Public Health, Icahn School of Medicine at Mount Sinai, New York, NY}
\affil[3]{Department of Biostatistics, Harvard T.H. Chan School of Public Health, Boston, MA}
\affil[4]{Department of Biostatistics, Columbia University Mailman School of Public Health, New York, NY}
\affil[$\star$]{Corresponding author. 722 W. 168th Street, New York, NY, 10032. mk3961@cumc.columbia.edu}

\begin{document}
%\sffamily

\maketitle

%\clearpage

\captionsetup[figure]{list=no}
%\captionsetup[table]{list=no}

\section*{Keywords:} Statistical learning, random seed, environmental mixtures, machine learning, Bayesian statistics, penalized regression.

% \section*{Word Count:} 2670
\clearpage
% \begin{linenumbers}

\section*{Abstract}

Statistical learning (SL) includes methods that extract knowledge from complex data. SL methods beyond generalized linear models, such as shrinkage methods or kernel smoothing methods, are being increasingly implemented in public health research and epidemiology because they can perform better in instances with complex or high-dimensional data---settings when traditional statistical methods fail. These novel methods, however, often include random sampling which may induce variability in results. Best practices in data science can help to ensure robustness. As a case study, we included four SL models that have been applied previously to analyze the relationship between environmental mixtures and health outcomes. We ran each model across 100 initializing values for random number generation, or ``seeds,'' and assessed variability in resulting estimation and inference. All methods exhibited some seed-dependent variability in results. The degree of variability differed across methods and exposure of interest. Any SL method reliant on a random seed will exhibit some degree of seed sensitivity. We recommend that researchers repeat their analysis with various seeds as a sensitivity analysis when implementing these methods to enhance interpretability and robustness of results.

\section*{Key Messages}

\begin{itemize}
    \item Statistical learning is increasingly useful for epidemiology applications as data dimensionality and complexity increase.
    \item Most statistical learning approaches incorporate random sampling. Defining a seed enables reproducibility.
    \item Findings may vary across seeds to different degrees, depending on the dataset and the chosen method.
    \item Sensitivity analyses should assess robustness of results to seed selection.
    \item If results are highly variable across seeds, a distribution of estimated effects across seeds should be presented.
\end{itemize}

\section*{Introduction}

As the data we use for epidemiologic studies become more complicated---with high-dimensional exposure and outcome spaces and increasing sample sizes---investigators are reaching for statistical learning (SL) tools, i.e., a set of tools for modeling and understanding complex datasets more suited to accommodate big data \cite{james2013introduction}. Indeed, epidemiologists are increasingly adopting SL methods to help answer research questions in public health. For example, epidemiologists have used clustering algorithms to determine the effect of PM$_{2.5}$ exposure across cities with different PM$_{2.5}$ composition \cite{kioumourtzoglou2016pm2}, penalized regression to identify chemical-specific independent associations between environmental contaminants and birth weight \cite{lenters2016prenatal}, and tree-based methods to assess potential interactions among air pollution toxins and their relationship with child cognitive skills \cite{stingone2017using}, among others. These approaches appear better equipped than traditional methods to accommodate numerous issues, such as model uncertainty, multi-collinearity, and multiple comparisons. 

Most well-established methods for SL incorporate random sampling, e.g., a tuning parameter that depends on a random division of the dataset or a random starting point to begin sampling \cite{bda3, murphy2012machine}. When applying SL to public health questions, understanding the role of random sampling in these methods and assessing the potential variability induced by this randomness can greatly increase confidence in the results.

Random dataset divisions are used by some SL methods to split the full dataset to a ``training'' dataset, to build the model, and a ``testing'' dataset to (a) measure the predictive ability of said model and/or (b) assist in estimating tuning parameters to maximize predictive accuracy. When no external testing dataset is available, cross-validation (CV) can be used in the full dataset; CV randomly splits the original dataset in two, builds the model on one set of the data (i.e., the training set), and evaluates its performance on the left-out set (i.e., the testing set). A common extension, $k$-fold CV, randomly partitions the original data into $k$ subsamples and repeats the training $k$ times using all but a single subsample, while testing the model on the held-out sample. The predicted values for each held-out sample are then compared to the observed ones from the same sample, for example by averaging the $k$ error estimates \cite{murphy2012machine}. CV works well when the objective of the statistical analysis is prediction; it may not perform as well to assess sensitivity and specificity or to evaluate effect estimates. Nonetheless, CV is often used for assessing goodness of fit and selecting tuning parameter values in epidemiologic models in lack of other options to optimize SL models for effect estimation.

The purpose of assessing model performance in a separate (``testing'') dataset is to avoid over-fitting and ensure generalizability of results. In CV, if each subsample is randomly drawn from the original dataset, then, on average, each should serve as a representative subsample. The key phrase here is ``on average''---it does not mean that any given subsample is representative of the complete sample. Researchers can enhance their confidence in the results simply by running a sensitivity analysis with a different seed, as the seed will determine the splitting. Similar results on tuning parameters chosen from different splits that arise from different seeds will increase confidence that the randomly chosen subsamples are representative of the whole and will strengthen conclusions regarding generalizability.

SL methods that rely on a Bayesian framework for statistical inference include---but are not limited to---Bayesian model averaging, e.g., to identify time windows of exposure to pollutants that produce adverse health effects \cite{dominici2008model}; Bayesian networks to account for uncertainty, as e.g., in infectious disease epidemiology \cite{lau2016bayesian}; and Bayesian hierarchical models, e.g., to investigate day-to-day changes in coarse particulate matter air pollution and cause-specific mortality \cite{chen2019associations}. These Bayesian methods involve a set seed for a different purpose than splitting the dataset to subsamples. Bayesian methods specify prior distributions that, when combined with a data likelihood, give rise to a posterior distribution of interest. Calculating the exact posterior distribution of most models is impossible or, at least, too computationally expensive to be reasonable. Instead, Markov Chain Monte Carlo (MCMC) is used to produce samples from a model's posterior distribution. The marginal distribution of a Markov chain that has converged to the stationary distribution can be used to make inferences that approximate results from the full posterior distribution. MCMC is inherently random and chains often begin from a random point, which is defined by the set seed. In theory, if the chain runs long enough to converge, then the initialization seed should not matter \cite{bda3}. In practice, convergence assessment is not always straight-forward. Although there are diagnostic tools that quantify the variation among multiple chains to assess whether the chains have converged, such as the Gelman-Rubin statistic \cite{bda3,gelman1992inference,cowles1996markov}, these are not always applied in epidemiologic studies. Similar results from a sensitivity analysis based on an initialization point assigned by a different seed will also increase confidence that the results represent the full posterior.

Assessing robustness of results from SL methods in epidemiology can be improved by sensitivity analyses. Here, we present a case study of analytic examples of sensitivity to seed selection across four SL models used in environmental epidemiology to assess exposure to environmental mixtures in health analyses, noting that this is but one example in public health. We discuss best practices to ensure robustness of results and the benefits of sensitivity analyses when utilizing SL to make inference in public health research.

\section*{Methods}

The aim of this analysis is to highlight the importance of incorporating seed sensitivity analyses into epidemiologic studies that employ SL methods, using as a case study an environmental epidemiology application. We employed four methods that depend on a seed, for either CV or as a random start for a sampling chain, to illustrate variability in results across different seeds. We availed ourselves of existing models from a study presenting multiple methods to assess exposure to multi-pollutant chemical mixtures in environmental epidemiology. Specifically, we used information on 18 persistent organic pollutants (POPs) and leukocyte telomere length (LTL) among 1,003 U.S. adults in the National Health and Nutrition Examination Survey (NHANES, 2001–2002) \cite{gibson2019overview}. The study population and exposure and outcome measurements have been described previously \cite{mitro2015cross, zipf2013health, cdc2002a, cdc2002b, akins1989estimation, cawthon2002telomere, lin2010analyses, needham2013socioeconomic}. LTL, the outcome variable in this study, refers to the length of the repetitive nucleotide sequence at the end of chromosomes that provides stability and allows for complete DNA replication in the ends \cite{blackburn2000telomere, greider1996telomere}. We  \textit{a priori} divided the 18 POP congeners, the exposure mixture, into three groups according to previous work: (1) eight non–dioxin-like PCBs, (2) two non-ortho PCBs (PCBs 126 and 169), and (3) POPs with moderate to high toxic equivalency factors (mono-ortho-substituted PCB 118, four dibenzo-furans, and three chlorinated dibenzo-p-dioxins), here referred to as mPFD \cite{mitro2015cross, gibson2019overview}. 

We built four models according to Gibson et al. \cite{gibson2019overview} including the same predictors of interest (POPs), covariates, and outcome (LTL). We included two penalized regression methods (lasso and group lasso) for variable selection that add a regularization term with an associated tuning parameter (often denoted using $\lambda$) to a regression model and chose the tuning parameter using ten-fold CV to control over-fitting. Lasso constrains the fit of a regression model with respect to the sum of the absolute values of the coefficients \cite{tibshirani96} and group lasso constrains the fit of the model with respect to the sum of the absolute values of \textit{a priori} defined groups \cite{friedman2001elements}. The regularization parameter $\lambda$ in each model controls the degree of shrinkage and is tuned to improve model fit and predictive capacity. In both of these methods, we only penalized the POP variables, but not the potential confounders included in the model.

A third model (weighted quantile sum regression, WQS) includes a single training and a hold-out set (testing set) to assess generalizability. WQS creates an empirically-weighted index of chemicals and includes this index as the exposure term in a regression model \cite{carrico15}. Here, CV is not used to tune the model; instead, the hold-out dataset is used to determine whether weights generalize from one random subsample to another. WQS also estimates a parameter that constrains each chemical weight to be between 0 and 1 and all weights to sum to 1. Important chemical components in the mixture are identified by comparing the weight for each component to a threshold parameter, $\tau$, chosen \textit{a priori}. Here we use $\tau=1/p$, where $p$ is the number of chemicals in the mixture, as has been previously suggested \cite{gibson2019overview,tanner2019repeated}. Because WQS is inherently one-directional, in that it tests only for mixture effects positively or negatively associated with a given outcome, we specified a positive unconstrained model.

The fourth model we included, Bayesian Kernel Machine Regression (BKMR), uses MCMC sampling (with a random seed for initialization) to estimate the posterior distributions of the model parameters by simulating realizations from these intractable posteriors \cite{bobb2014bayesian}. BKMR estimates chemical-specific exposure--response functions, detects potential interactions among mixture members, and estimates the overall mixture effect. Its variable selection feature also outputs posterior inclusion probabilities (PIPs). PIP values range between 0 and 1 and their magnitude indicates relative variable importance \cite{bobb2018statistical}. For this analysis, we performed hierarchical variable selection for the three pre-defined groups of congeners. This feature outputs both group PIPs and congener-specific conditional PIPs, i.e., the relative importance of a congener given that the group that contains that congener is important. 

We ran lasso, group lasso, WQS, and BKMR 100 times each, using a different seed each time. We measured variability in lasso and group lasso results using the proportion of nonzero beta coefficients and the median and inter-quartile range (IQR) of beta coefficients for each of the POPs across seeds. We measured variability in WQS results using (a) the proportion of weights above our chosen threshold ($1/p$) and the median and IQR of weights for each exposure of interest across seeds, and (b) pooling the WQS index coefficients across seeds using Rubin's rule \cite{rubin1987multiple,barnard1999miscellanea}. We measured variability in BKMR results by visualizing the full exposure--response curves and using the median and IQR of the posterior inclusion probabilities (PIP) for each of the POPs across seeds. All analyses were conducted in R version 3.6.1 \cite{rrr}, and all code to recreate results and figures is available online at \url{https://github.com/yanellinunez/Commentary-to-mixture-methods-paper}.

\section*{Results}

As expected, for all methods, we observed some variability in the results based on seed number. 

\subsection*{Lasso and Group Lasso} 
\begin{figure}[htbp]
    \centering
\includegraphics[scale=0.75]{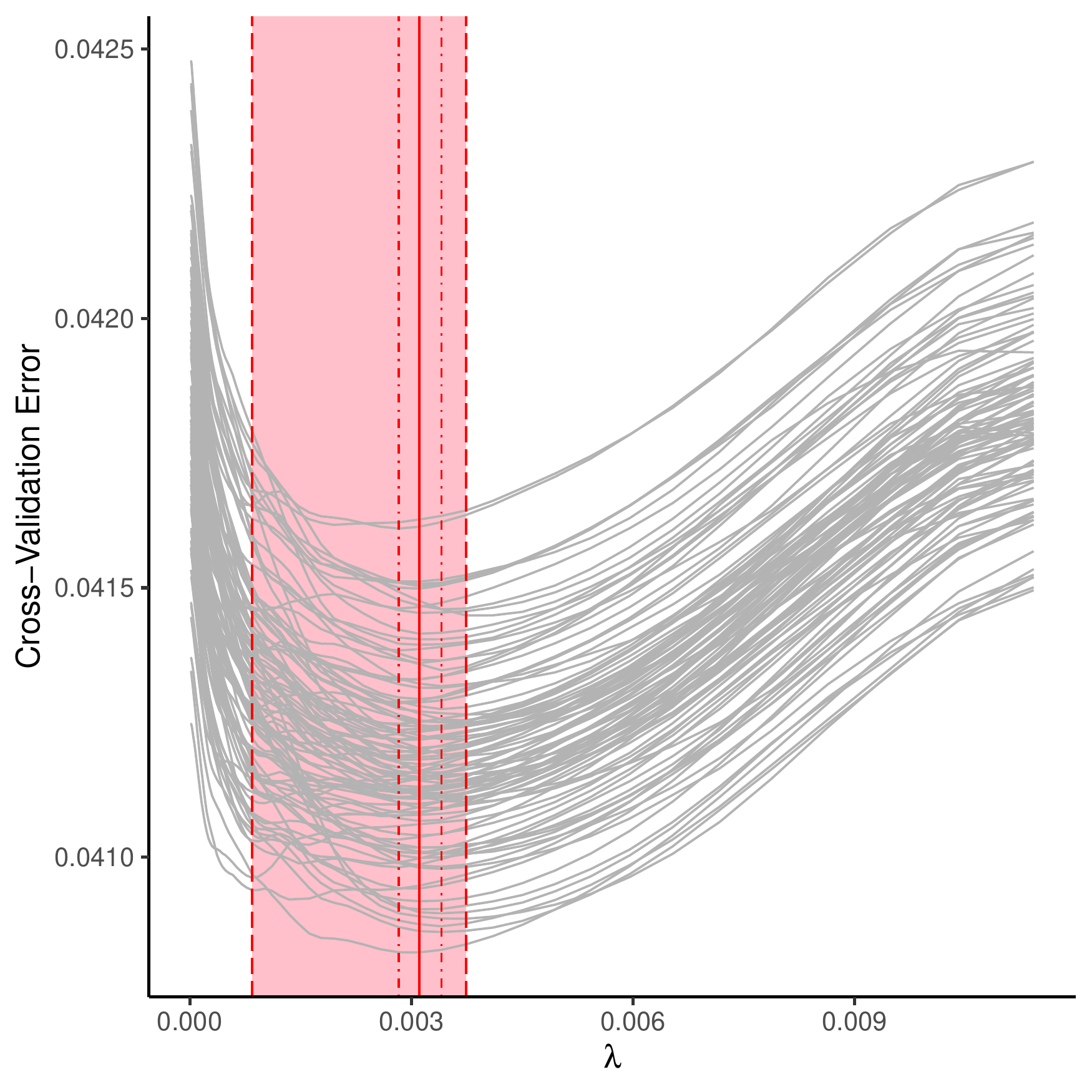}
   \caption{Cross-validation curves for lasso models over 100 seeds. Grey curves represent results obtained from each of the 100 seeds. The vertical red line indicates the median of the best fit $\lambda$ values. The inner dot-dashed lines represent the inter-quartile range of the best fit $\lambda$ values. The outer dashed lines represent the minimum and maximum of the best fit $\lambda$ values. All best fit $\lambda$ values fell inside the shaded area.}
    \label{fig:lasso_cv}
\end{figure}

Lasso models varied slightly across seeds, beginning with different $\lambda$ values chosen as the optimal tuning parameter (Figure~\ref{fig:lasso_cv}). While there was little difference in predictive accuracy across seeds (total CV error range = 0.041--0.042), the chosen $\lambda$ value based on CV affected the number of coefficients that were pushed to zero. The smallest $\lambda$ value selected, chosen by two of the 100 seeds, retained nine of the 18 congeners in the model. The largest $\lambda$ value selected, chosen by eleven of the 100 seeds, retained four congeners in the model. Sixty percent of all chosen $\lambda$ values retained five congeners. However, independent of seed, nine out of the 18 congeners consistently had beta coefficients of zero (dioxin 1,2,3,4,6,7,8-hpcdd, furans 1,2,3,4,7,8-hxcdf and 1,2,3,6,7,8-hxcdf, and PCBs 74, 138, 153, 170, 187, and 194); and four out of the 18 congeners (PCB 99, 118, 126, and furan 2,3,4,7,8-pncdf) consistently had nonzero beta coefficients. Four of the five other congeners, dioxins 1,2,3,4,6,7,8,9-ocdd and 1,2,3,6,7,8-hxcdd and PCBs 180 and 169 had nonzero coefficients in 10\% or less of the cases. Only one congener, furan 1,2,3,4,6,7,8-hxcdf, had less consistency, with nonzero beta coefficients in 71\% of the seeds (Figure~\ref{fig:grlasso_lasso}). 

In group lasso, the non-dioxin-like PCBs had the most variability across seeds relative to the other groups. This congener group had nonzero beta coefficients in 75\% of the seeds. In contrast, the mPFD and non-ortho PCBs consistently had nonzero beta coefficients. The non-dioxin-like PCBs also showed the widest range of beta values for a given congener across seeds, particularly PCBs 180 and 153. The non-ortho PCBs had consistent coefficient values across seeds (Figure~\ref{fig:grlasso_lasso}). 

\subsection*{WQS Regression} 

\begin{figure}[htbp]
    \centering
\includegraphics[width=.95\textwidth]{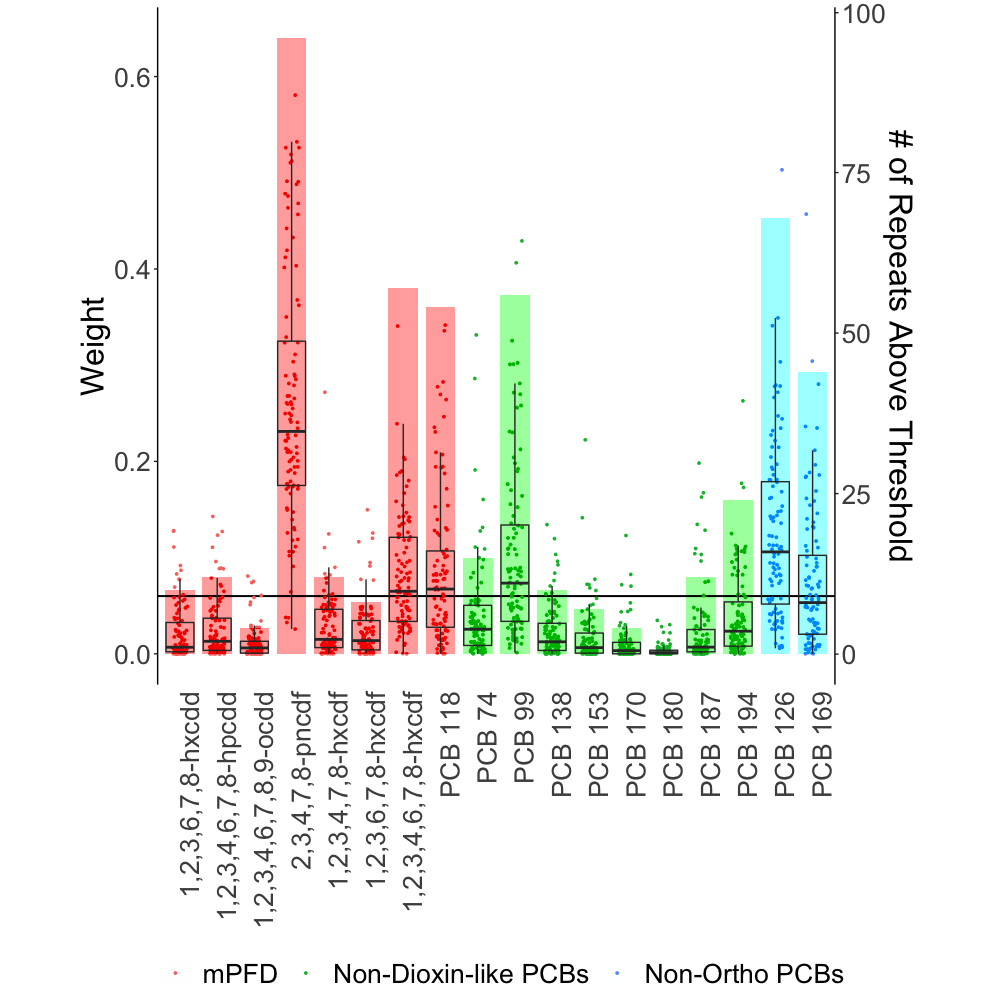}
   \caption{WQS\footnote{WQS: Weighted Quantile Sums}-estimated weights over 100 seeds. Bars correspond to the right axis and indicate the number of times a congener had a weight value above the threshold, which was calculated as $1/p$ (1/18 = 0.05, horizontal line) out of the 100 seeds. Data points and boxplots correspond to the left axis. The data points are the congeners' weights in each of the seeds. The boxplots show the median and interquartile range for the weights of each congener across seeds.}
    \label{fig:wqs_summary}
\end{figure}

\begin{figure}[htbp]
    \centering
\includegraphics[width=.95\textwidth]{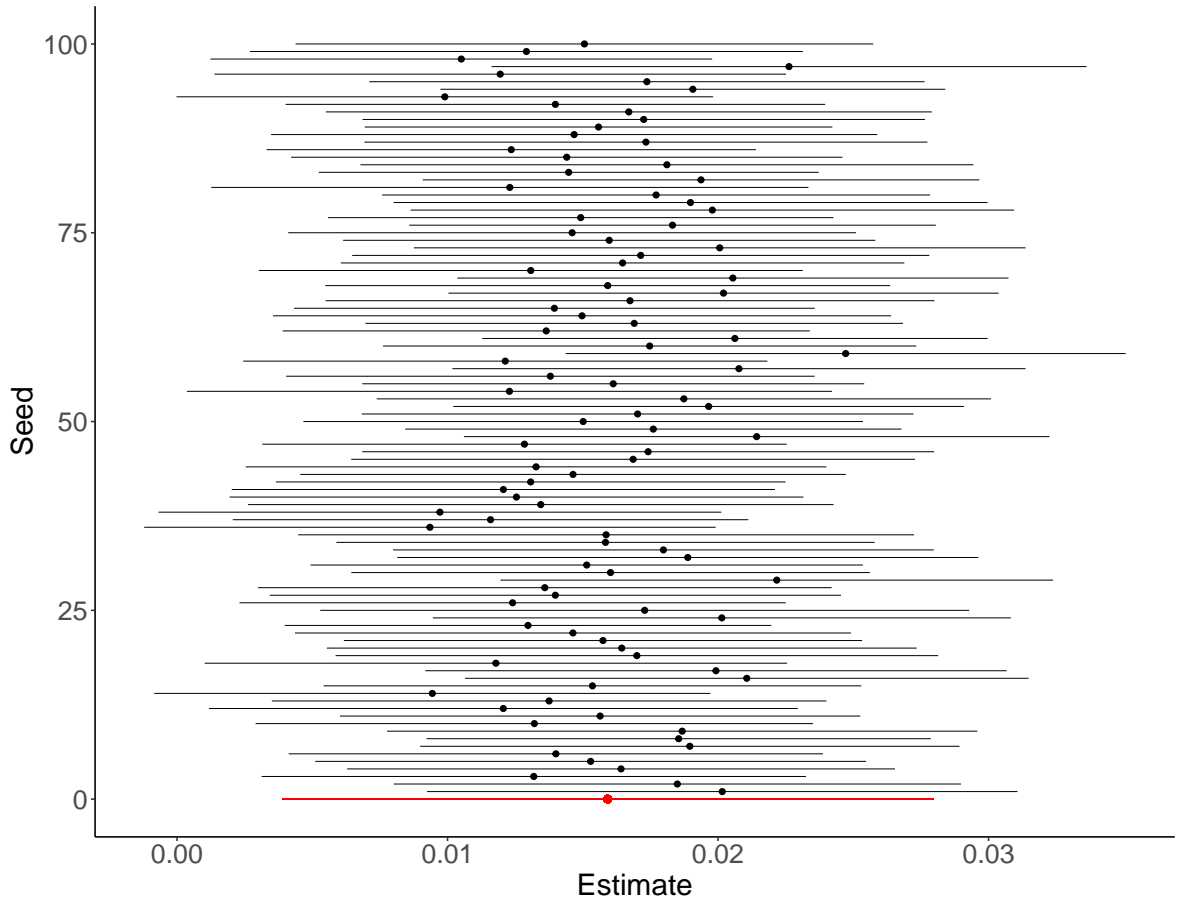}
   \caption{WQS\footnote{WQS: Weighted Quantile Sum} index estimates over 100 different seeds. Black points and black lines represent estimates for each of the 100 seeds and the 95\% confidence intervals, respectively. The red point represents the pooled estimate across all seeds and its 95\% confidence interval.}
    \label{fig:wqs_index}
\end{figure}

In WQS, weight values varied to a degree across seeds, which resulted in the number of weight values above the threshold $\tau$ and the congener order of importance varying from seed to seed. Out of the hundred seeds, furan 2,3,4,7-8-pncdf and PCB 126 had weight values above the threshold 96\% and 68\% of time, respectively. These two congeners also had the largest weight in most instances---in 49 and 15 of the seeds, respectively. PCB 99, PCB 118, and furan 1,2,3,4,6,7,8-hxcdf had weights above the threshold in approximately 50\% of the seeds and presented the largest weight value in 12, 8, and 4 instances, respectively. The remaining congeners showed weights above the threshold in fewer than 50\% of the seeds (Figure~\ref{fig:wqs_summary}). The regression coefficient for the pooled WQS index across seeds was significantly positive. The confidence intervals around the effect estimate for the WQS index did not include the null value for 97\% of the seeds based on the hold-out data (Figure~\ref{fig:wqs_index}). 

\subsection*{BKMR}

\begin{figure}[htbp]
    \centering
\includegraphics[width=.95\textwidth]{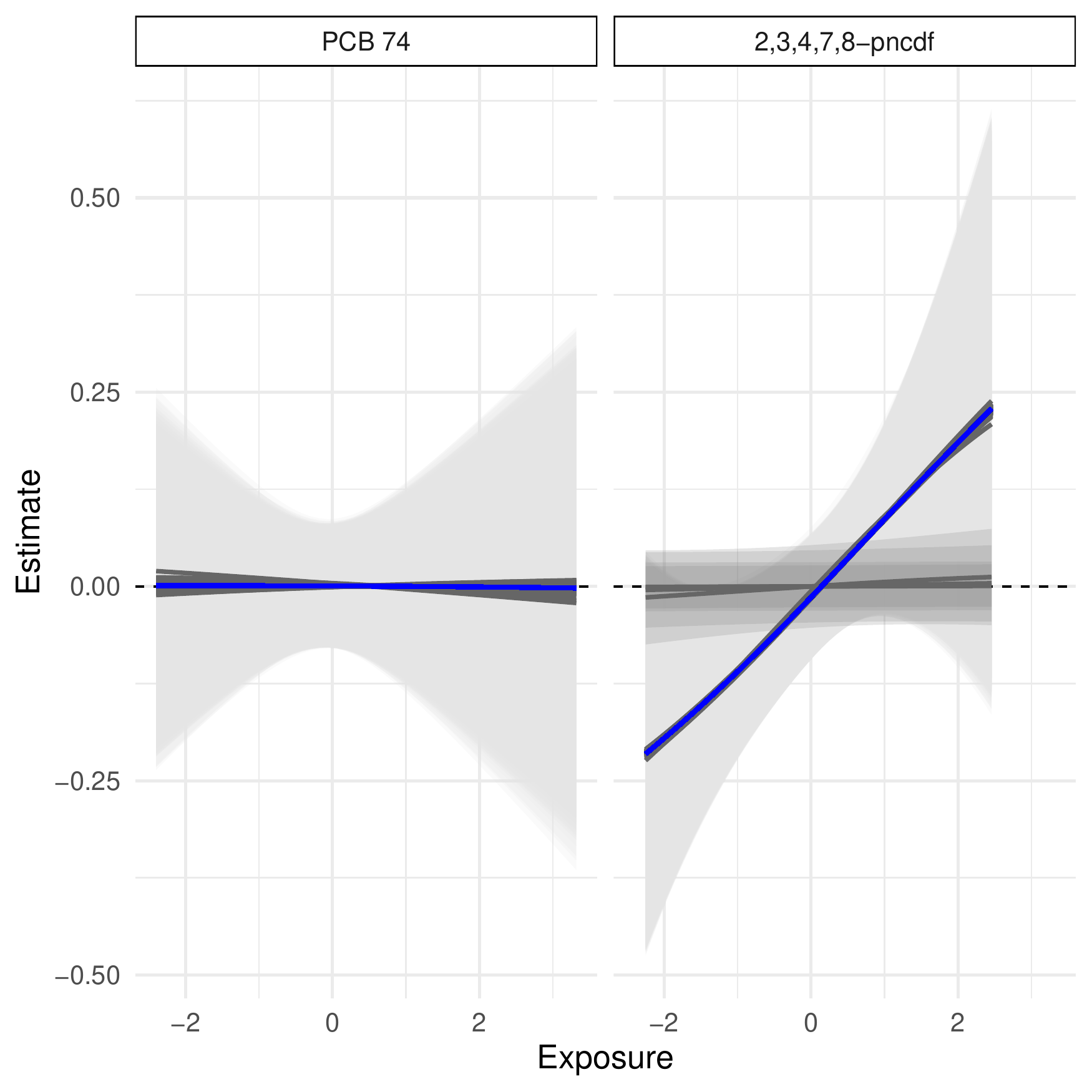}
   \caption{Congener-specific effect estimates for two mixture members, PCB 74 (non-dioxin-like PCBs) and furan 2,3,4,7,8-pncdf (mPFD) estimated by BKMR\footnote{BKMR: Bayesian Kernel Machine Regression} over 100 MCMC\footnote{MCMC: Markov Chain Monte Carlo} seeds. Grey lines represent exposure--response functions for each of the 100 seeds. The blue line indicates the median across all seeds.}
    \label{fig:bkmr}
\end{figure}

For MCMC seeds, the order of PIPs for the congener groups and the highest conditional PIP within each group (furan 2,3,4,7,8-pncdf for mPFD, PCB 126 for non-ortho PCBs, and PCB 170 for non-dioxin-like PCBs) did not change based on seed selection (Table~\ref{table:bkmr}). Direction and magnitude of exposure-response functions for individual POPs across seeds were largely similar, with notable deviations for PCBs 99, 118, 126, 169, and 180, and furan 2,3,4,7,8-pncdf. For these six congeners, four of the hundred seeds produced null exposure--response curves with no suggestion of a positive or negative association with LTL. For all other seeds, the credible intervals around furan 2,3,4,7,8-pncdf did not include the null value at low levels of the exposure--response curve (Figure~\ref{fig:bkmr}) and we observed suggestive evidence of associations with PCBs 126 and 169. For 96 of 100 seeds, we observed an overall mixture effect, with null results for the same four seeds.

\section*{Discussion}

Our goal was to show the value of incorporating multiple seeds when applying SL methods that involve a random process in epidemiologic studies. We used an environmental mixtures example as a case study and ran four methods over one hundred different seeds, to illustrate the benefit of sensitivity analyses. Obtaining results across multiple seeds increases the generalizability of the results and the conclusions drawn. We showed that although setting a specific seed may ensure the reproducibility of an analysis, it does not guarantee generalizability and robustness, highlighting the benefit of estimating parameters for a given SL model using multiple seeds.

For methods applying some form of CV to select the optimal tuning parameter for the model, sensitivity to seed gives rise to variability in the choice of this parameter (Figure~\ref{fig:lasso_cv}). While there may be no practical significance in epidemiologic research between two $\lambda$ values in terms of predictive accuracy measured by CV error, two similar $\lambda$ values may result in substantial differences in variable selection. In our case study, the number of retained congeners varied between four and nine, depending on the optimal $\lambda$ value selected by CV error across seeds, with five congeners selected in 60\% of the seeds. This could result in quite different conclusions about the potential toxicity of the examined congeners.

Variability in $\lambda$, thus, may affect the stability of beta coefficients across seeds (Figure~\ref{fig:grlasso_lasso}). In the case of lasso, nine beta coefficients were consistently pushed to zero and four consistently push to nonzero, across all seeds. However, one congener (furan 1,2,3,6,7,8-hxcdf) had a positive beta coefficient for 71 out of 100 seeds and four other congeners (dioxins 1,2,3,4,6,7,8,9-ocdd and 1,2,3,6,7,8-hxcdd and PCBs 180 and 169) had nonzero coefficients less than 10\% of the time. Thus, if this analysis were run with a single seed, it would have been possible to find and report that furan 1,2,3,6,7,8-hxcdf was not predictive of the outcome or that PCB 180 was. Depending on the context, the consequences of ruling out a potential association with one POP or drawing undue attention to another may be trivial or not. In the case of group lasso, all congeners within \textit{a priori}-defined groups were penalized together. Congeners in the mPFD and non-ortho PCB groups consistently had nonzero beta coefficients over the hundred seeds. Congeners in the non-dioxin-like PCBs group had nonzero beta coefficients in 75 seeds, but had beta coefficients of zero in the remaining 25 seeds. Again, if the analysis were run with only one seed, the predictive power of non-dioxin-like PCBs may have been missed. 

In the case of WQS, the dataset is randomly partitioned into a training and a testing set. The weights for each congener are then calculated from the training set and subsequently used to estimate the mixture index with the testing set. Thus, the results can be strongly influenced by the initial random partition of the data and may be unstable, especially when the dataset is small. Our analysis, with a sample size of 1,003, showed that the number of congeners with a weight above the threshold and the magnitude of given weights varied across seeds. Tanner et al. \cite{tanner2020early, tanner2019repeated} address this issue by using a repeated hold-out validation. This process repeats the analysis one hundred times over randomly selected seeds, similar to what we did here, then reports the mean and confidence intervals for the sampling distribution of the weights instead of a single estimate from a single seed, enhancing generalizability. Code for this technique is also publicly available (\url{https://github.com/evamtanner/Repeated_Holdout_WQS}). Although we found some variability in the estimated congener-specific weights, the conclusion that this mixture has an overall harmful effect on LTL was robust to seed selection.  

BKMR results varied little based on the selected seeds, but four of the hundred chosen MCMC seeds failed to identify any non-null univariate exposure--response functions or an overall mixture effect. Since seed should not influence results if the Markov chain has converged to its stationary distribution, we believe that the runs based on these four seeds did not converge. It is likely that the chain became ``stuck'' in an area with a local optimum and 100,000 iterations were not sufficient to examine the full distribution. Further inspection of convergence criteria, however, provided no indication of failure to converge, supporting our recommendation of sensitivity analyses with multiple seeds or the use of convergence diagnostics, such as the Gelman-Rubin statistic.

Recommendations of this sort exist already in various subject areas. In machine learning, re-fitting a model, such as a penalized regression or a tree-based method, on bootstrapped samples of the data has been recommended \cite{chatterjee2011bootstrapping, friedman2001elements}. This is another means of capturing variability attributed here to seed selection. A single regression tree, for example, is not very stable; a different seed will divide the data into different training and testing sets that may result in vastly different trees. A random forest, which trains multiple trees on bootstrapped samples of observations and uses only a random subset of variables at each split, is much more robust. The method we used to pool WQS coefficients across seeds, Rubin's rule, is commonly implemented in stochastic missing-data imputation to combine variance between and within multiple imputations \cite{rubin1987multiple, barnard1999miscellanea}. For Bayesian MCMC, it is often recommended to simulate multiple chains and then check that they have converged to the same distribution, i.e., ``mixed'' \cite{levin2017markov}. In this reflection on seed selection in SL, we aim to connect applied epidemiologic research to these avenues.

A degree of variability across seeds is expected due to the intrinsic randomness associated with the methods at hand. However, seed sensitivity is not specific to these four methods; any SL tool may be susceptible. Nor is the generalizability and robustness of results only method-dependent---factors such as sample size or data heterogeneity should also be considered. Finally, we note that many SL methods, including lasso and group lasso, were developed to improve predictive accuracy. Their ability to accommodate complex and high-dimensional datasets make them increasingly appealing tools for use in epidemiologic analyses; application, nonetheless, to estimate health effects should be performed with caution.

\section*{Conclusion}

Running the same analysis under different seeds provides a better understanding of the results. Thus, we recommend that epidemiologists employing SL methods run models that involve a random component with multiple seeds as best practice. Results across seeds should not be used to select a seed number, but instead as sensitivity analysis to assess the robustness of the results and enhance generalizability of study findings. A randomly selected seed should be used for the main analysis, and results across seeds should be included as supplementary material. When results across seeds vary greatly, researchers should consider reporting averages and inter-quartile ranges rather than an estimate from a single seed. 

\section*{Funding}

This work was partially supported by the National Institutes of Environmental Health (NIEHS) PRIME R01 grants [ES028805, ES028811, and ES028800] and Centers [P30 ES009089, P30 ES000002, P30 ES023515, and U2C ES026555], as well as F31 ES030263 and R01 ES028805-S1.

\section*{Conflicts of Interest} None declared.

\clearpage
\bibliographystyle{ieeetr}
\bibliography{main}
% \end{linenumbers}

\clearpage
\section*{Tables}

\floatsetup[table]{capposition=top}
\begin{table}[htbp] \centering 
\caption{Posterior Inclusion Probabilities (PIPs) for Persistent Organic Pollutants groups and conditional PIPs for individual congeners across 100 Markov Chain Monte Carlo seeds.}
  \label{table:bkmr} 
\begin{tabular}{@{\extracolsep{5pt}} llccc} 
\hline \hline 
\multirow{2}{*}{Group} & & \multicolumn{3}{c}{Posterior Inclusion Probabilities} \\ 
& & Minimum & Median & Maximum \\ 
\hline \\[-1.8ex] 
mPFD && 0.65 & 0.86 & 0.89  \\ 
Non-Ortho PCBs && 0.61 & 0.67 & 0.76 \\ 
Non-Dioxin-like PCBs  && 0.41 & 0.46 & 0.68 \\ 
\hline \\[-1.8ex] 
\multirow{3}{*}{Group} & \multirow{3}{*}{Congener} & \multicolumn{3}{c}{Conditional Posterior} \\ 
& & \multicolumn{3}{c}{Inclusion Probabilities} \\ 
& & Minimum & Median & Maximum \\ 
\hline \\[-1.8ex] 
\multirow{8}{*}{mPFD} & 1,2,3,6,7,8-hxcdd & 0.01 & 0.01 & 0.12 \\ 
& 1,2,3,4,6,7,8-hpcdd & 0.01 & 0.01 & 0.13 \\ 
& 1,2,3,4,6,7,8,9-ocdd & 0.004 & 0.01 & 0.12 \\ 
& 2,3,4,7,8-pncdf & 0.12 & 0.86 & 0.89 \\ 
& 1,2,3,4,7,8-hxcdf & 0.01 & 0.02 & 0.13 \\ 
& 1,2,3,6,7,8-hxcdf & 0.01 & 0.02 & 0.13 \\ 
& 1,2,3,4,6,7,8-hxcdf & 0.01 & 0.02 & 0.14  \\
& PCB 118 & 0.05 & 0.06 & 0.14  \\
\hdashline \\[-2ex] 
\multirow{2}{*}{Non-Ortho PCBs} & PCB 126 & 0.51 & 0.65 & 0.68 \\
& PCB 169 & 0.32 & 0.35 & 0.49  \\ 
\hdashline \\[-2ex] 
\multirow{8}{*}{Non-Dioxin-like PCBs} & PCB 74 & 0.08 & 0.10 & 0.13  \\ 
& PCB 99 & 0.11 & 0.13 & 0.16 \\ 
& PCB 138 & 0.10 & 0.12 & 0.14  \\ 
& PCB 153 & 0.12 & 0.14 & 0.16 \\ 
& PCB 170 & 0.12 & 0.17 & 0.21 \\ 
& PCB 180 & 0.11 & 0.13 & 0.17  \\ 
& PCB 187 & 0.08 & 0.09 & 0.13 \\ 
& PCB 194 & 0.09 & 0.10 & 0.12 \\ 
\hline \\[-2.8ex]
\end{tabular} 
\end{table} 

\clearpage

\section*{Figures}

\clearpage
\begin{landscape}
\captionsetup[subfigure]{position=top}
\begin{figure}[htbp]
    \centering
    \subfloat[Lasso]{%
   \includegraphics[width=.45\textwidth]{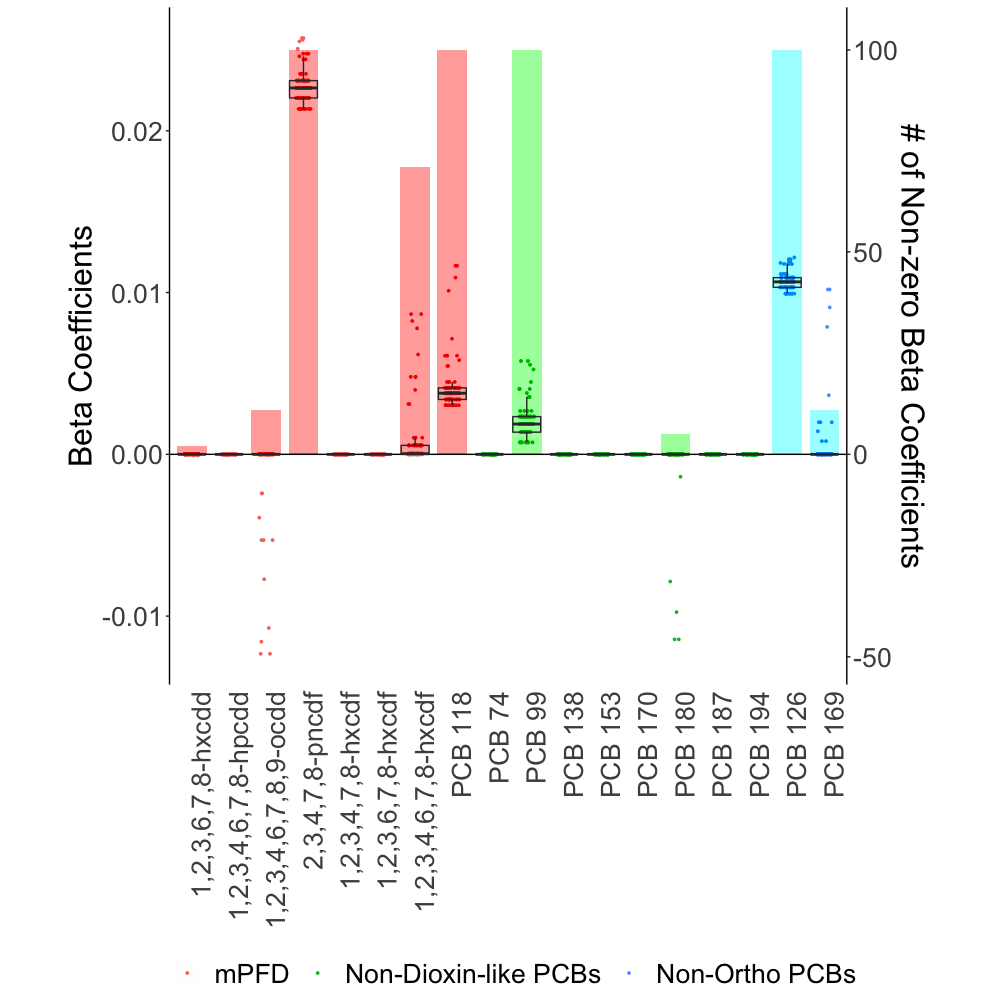}%
 }
    \qquad
 \subfloat[Group Lasso]{%
 \includegraphics[width=.45\textwidth]{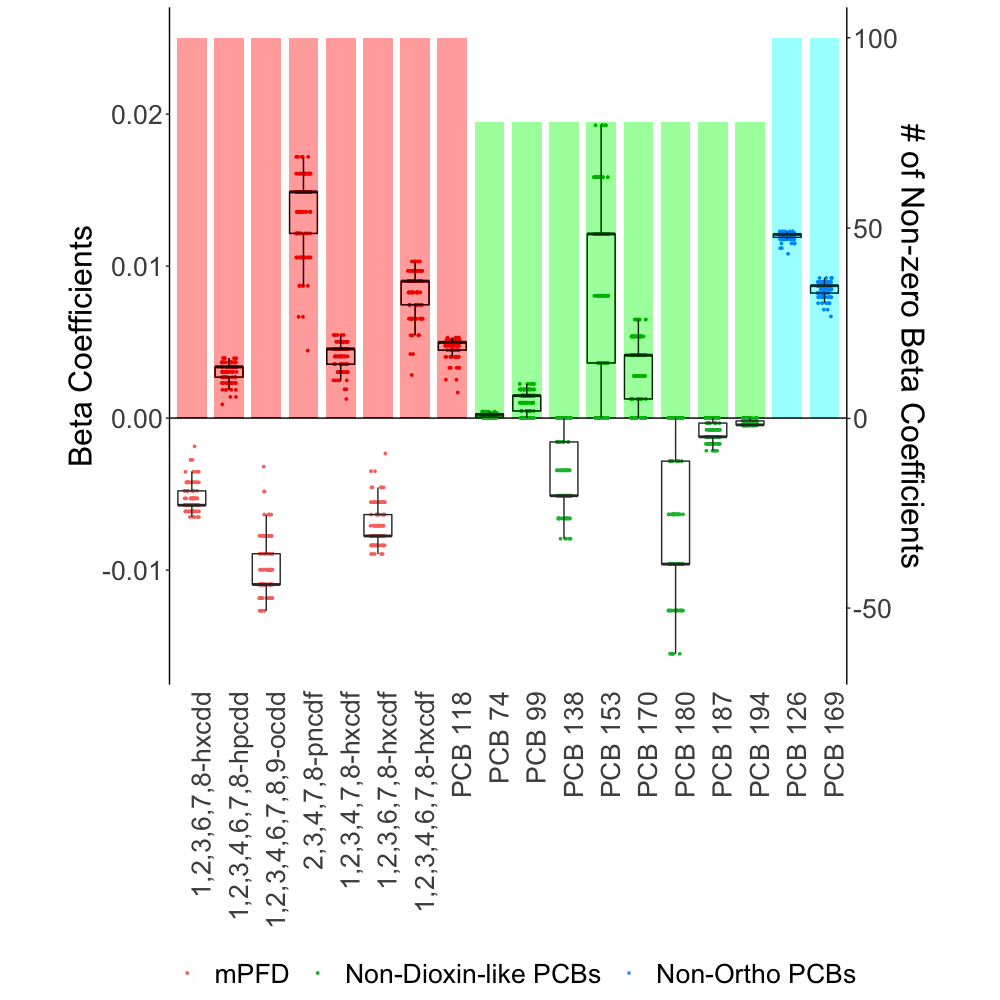}%
 }%
   \caption{\textbf{(a)} Lasso and \textbf{(b)} Group Lasso beta coefficients over 100 seeds. Bars correspond to the right axis and indicate the number of times a congener had a nonzero beta coefficient across all seeds. Data points and boxplots correspond to the left axis. The data points are the congeners' beta coefficients in each of the seeds. Boxplots show the median and interquartile range for the beta coefficients of the given congener.}
    \label{fig:grlasso_lasso}
\end{figure}
\end{landscape}
\clearpage

\clearpage

\end{document}